\documentclass[sn-mathphys,Numbered]{sn-jnl}

\usepackage{graphicx}%
\usepackage{multirow}%
\usepackage{amsmath,amssymb,amsfonts}%
\usepackage{amsthm}%
\usepackage{mathrsfs}%
\usepackage[title]{appendix}%
\usepackage{xcolor}%
\usepackage{textcomp}%
\usepackage{manyfoot}%
\usepackage{booktabs}%
\usepackage{algorithm}%
\usepackage{algorithmicx}%
\usepackage{algpseudocode}%
\usepackage{listings}%
\usepackage{color,soul}%

\theoremstyle{thmstyleone}%
\theoremstyle{thmstyletwo}%

\theoremstyle{thmstylethree}%

\begin{document}

\title[Ultrafast modulation of guided-mode resonance in a nonlinear silicon nitride grating]{Ultrafast modulation of guided-mode resonance in a nonlinear silicon nitride grating}


\author*[1,3]{\fnm{Subhajit} \sur{Bej}}\email{subhajit.bej@gmail.com}


\author[2]{\fnm{Nikolai} \sur{Tkachenko}}\email{nikolai.tkachenko@tuni.fi}
\author[1]{\fnm{Robert} \sur{Fickler}}\email{robert.fickler@tuni.fi}
\author[1]{\fnm{Tapio} \sur{Niemi}}\email{tapio.niemi@tuni.fi}

\affil*[1]{\orgdiv{Photonics Laboratory, Physics Unit}, \orgname{Tampere University}}

\affil[2]{\orgdiv{Chemistry and Advanced Materials Group, Faculty of Engineering and Natural Sciences}, \orgname{Tampere University}}

\affil[3]{\orgname{Dispelix Oy}, \orgaddress{\street{Metsänneidonkuja 10}, \city{Espoo}, \postcode{02130}, \country{Finland}}}


\abstract{

Nonlinear optical interactions in nanostructures are subjects of both fundamental research and practical applications. The optical Kerr effect, a third-order nonlinear optical phenomenon, induces anisotropic changes in the refractive index of a material under intense laser illumination. This effect is inherently weak, limiting its practical use in free-space nanophotonics. In this work, we demonstrate an enhancement of the optical Kerr effect by more than three orders of magnitude using guided-mode resonance in a dielectric diffraction grating with an overall thickness of less than 300 nm. Our study involves the design, fabrication, and measurement of the transmittance of a resonant silicon nitride crossed grating illuminated with short light pulses. We observe spectral resonance shifts with varying pulse power, leading to more than fifty percent modulation of the resonance magnitude and enabling transitions between `Off' and `On' states of transmission. We also demonstrate the capability of the grating for dynamic pulse shaping. The results presented here reveal promising directions for developing advanced all-optical devices based on free-space nanophotonics.
}

\keywords{Guided-mode resonance, optical Kerr effect, self-phase modulation, all-optical modulator, dynamic pulse shaping, diffraction gratings}



\maketitle

\section*{Introduction}\label{sec1}
Since the pioneering experiments on second-harmonic generation (SHG) \cite{franken1961}, and optical wave mixing \cite{armstrong1962}, the field of nonlinear optics has grown substantially. Recent research has focused on integrating novel nonlinear optical materials into miniaturized devices for practical applications \cite{stegeman1985,lapine2014}. This progress has been driven by advancements in nanofabrication and materials science, which have enabled the development of innovative materials, such as metal-dielectric nanocomposites \cite{boyd1996} and epsilon-near-zero materials \cite{reshef2019}, both recognized for their remarkable nonlinear optical responses.

The optical Kerr effect (OKE), a third-order nonlinear phenomenon, induces anisotropic refractive index changes in materials under intense laser illumination \cite{boyd2008}. This effect has been harnessed in integrated optics to develop various components, including all-optical switches, logic gates, and bistable optical elements \cite{stegeman1988,fortenberry1986,pardo1987,assanto1986,jensen1982,pu2010,radic1994,radic1995,lattes1983,vincent1985,hegde2005}. OKE has also been explored in free-space diffractive elements \cite{manela2007}. While free-space structures are advantageous for large-scale parallel all-optical signal processing, they encounter challenges due to short light-matter interaction lengths, which limit volumetric nonlinear optical effects.

To address these limitations, researchers have concentrated on synthesizing materials with stronger nonlinear optical responses and leveraging strong electromagnetic field confinement in optically thin materials. These objectives can be achieved through advanced material engineering and by exploiting resonances in sub-wavelength diffractive structures, respectively \cite{lapine2014}. Although plasmonic structures can enhance nonlinear signals \cite{kauranen2012}, their high Ohmic losses limit their practical applications. In contrast, all-dielectric nanophotonics presents a promising alternative, offering negligible losses and high damage thresholds \cite{kivshar2018,sain2019,smirnova2016}.

Among all-dielectric structures, guided-mode resonant (GMR) gratings stand out for their high-Q resonance, strong field confinement, and tunability \cite{quaranta2018}. Initially introduced by Mashev and Popov \cite{mashev1984,mashev1985} and further explored by Golubenko et al. \cite{golubenko1985}, GMR gratings have been proven to be useful in laser resonators, narrow-band filters, and selective mirrors \cite{avrutskii1986,golubenko1986}. While early studies focused on static devices, dynamic tuning via quadratic electro-optic effects in GMR gratings was later introduced by Sharon et al. \cite{sharon1996}, leading to further exploration of mechanical and nonlinear tuning mechanisms. Numerical studies on GMR-based optical switching and bistability \cite{boye1999,vincent1985,ngo2009,kang2017} and experimental studies on tunable GMRs using slow nonlinear effects in Azobenzene dye \cite{dobbs2006,yang2008} have expanded understanding of the potential of these structures.

In this study, we present the first experimental demonstration of ultrafast nonlinear optical modulation in a guided-mode resonant (GMR) grating. Our results show over 50\% modulation of the resonance magnitude across the spectrum, driven by an optical Kerr effect (OKE)-induced shift in the resonance peak, which enables dynamic switching between 'On' and 'Off' transmission states of the grating. For this experiment, we fabricated a grating consisting of square-shaped pillars arranged in a square lattice, using low-loss SiN$_x$ material, which exhibits moderately high intrinsic self-defocusing nonlinearity within the relevant spectral range. The experimental findings are supported by comprehensive numerical simulations.

We also employ a semi-analytical approach to elucidate the mechanisms underlying the waveguide-enhanced nonlinear optical effects and the associated resonance peak shift. To further understand the origins of the GMR-enhanced OKE, we visualize the field and permittivity distributions within the structure. Additionally, we demonstrate the potential of this grating structure for all-optical pulse shaping and nonlinear material characterization.

\section*{Theory and Results}\label{sec2}
\subsection*{Semi-analytical model}

We begin by deriving a semi-analytical model to elucidate the nonlinear optical modulation mechanism in the guided-mode resonance (GMR) grating. A schematic representation of the grating is depicted in Fig.~1(a), featuring a crossed surface-relief binary grating on top of a waveguide layer. The medium of incidence and the substrate regions consist of homogeneous materials with refractive indices $n_1$ (air) and $n_3$ (fused silica), respectively. Both the grating and the waveguide are fabricated from silicon nitride (SiNx) material with a refractive index of $n_\mathrm{wg}$. The grating exhibits periodicities $d_x$ and $d_y$, and fill factors $f_x$ and $f_y$ along the x- and y-axes, respectively. The top grating diffracts the incident light, guiding it into discrete modes within the planar waveguide underneath. The interplay between the diffraction and the waveguide modes induces resonant behavior, characterized by rapid variations in the amplitudes of propagating waves \cite{quaranta2018}.

\begin{figure}[H]
\centering
\includegraphics[width=\textwidth]{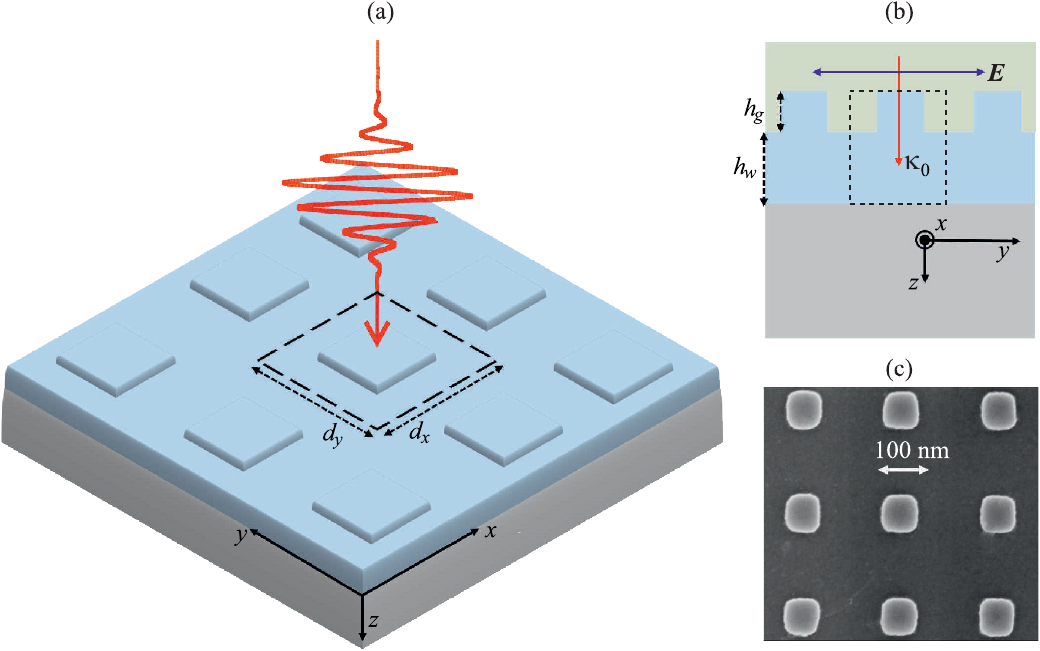}
\caption{Geometry of the GMR grating. (a) Three-dimensional (3D) perspective of an optical pulse incident on the grating from air. One grating period is marked with the dotted rectangle. (b) Side view showing that the wave vector $\mathbf{\kappa}_0$ is along the positive $z$-axis, and the electric field ($\mathbf{E}$) is polarized along $y$-axis. (c) A scanning electron micrograph (top-view) of the fabricated GMR grating.} \label{fig1}
\end{figure}

The spectral positions of the resonance maxima (or minima) of a GMR, can be qualitatively explained by examining the waveguide modes of the unperturbed slab waveguide with a thickness denoted as $h_{\mathrm w}$ \cite{amedalor2023}. For such a slab waveguide, modes can be categorized as TE (TM), where the $\mathbf{E}$-field ($\mathbf{H}$-field) is perpendicular to the direction of propagation \cite{pollock2003}. The eigenvalue equations governing TE and TM polarized slab waveguide modes can be expressed as follows:
\begin{equation}
\tan{(\kappa_p h_{\mathrm w})}=\frac{\kappa_i(\gamma_p+\delta_p)}{\kappa_p^2-\gamma_p\delta_p},\hspace{20mm}(\mathrm{TE})
\label{eq:1}
\end{equation}
\begin{equation}
\tan{(\kappa_p h_{\mathrm w})}=n_{wg}^{2}\kappa_p\frac{(n_3^2\gamma_p+n_1^2\delta_p)}{(n_1^2n_3^2\kappa_p^2-n_{wg}^4\gamma_p\delta_p)}.\hspace{20mm}(\mathrm{TM})
\label{eq:2}
\end{equation}
In Eqs.~\eqref{eq:1} and \eqref{eq:2}, $\kappa_p = (n_\mathrm{wg}^2\kappa_0^2 - \beta_p^2)^{1/2}$, $\gamma_p = (\beta_p^2 - n_1^2\kappa_0^2)^{1/2}$, and $\delta_p = (\beta_p^2 - n_3^2\kappa_0^2)^{1/2}$ represent the transverse components of the propagation vectors inside the waveguide layer, the superstrate, and the substrate, respectively. Here, $\kappa_0 = 2\pi/\lambda_0$ is the wave number, and $\lambda_0$ is the wavelength in air. The propagation constants ($\beta_p$) governing the slab waveguide modes can be determined by solving Eqs.\eqref{eq:1} and \eqref{eq:2}. The spectral dispersion of the effective indices of the slab waveguide modes, $n_{\mathrm{eff}}^{\mathrm{p}}=\beta_p/\kappa_0$, for a waveguide thickness of $h_{\mathrm w}=245$ nm is minimal, as illustrated in Fig.~S6 of the supplementary information. The guiding layer is surrounded by a fused silica substrate with a refractive index of $n_3=1.45$ and a cover material with a refractive index of $n_1=1.0$. This slab waveguide can support one TE mode (fundamental TE or TE$_0$) and one TM mode (fundamental TM or TM$_0$) with effective mode indices of $n_{\mathrm{eff}}^{\mathrm{TE_0}}=1.76$ and $n_{\mathrm{eff}}^{\mathrm{TM_0}}=1.66$, respectively. Resonance anomalies are observed when the diffraction of the grating aligns with the tangential wave number of the incident wave, matching the propagation constant of a slab waveguide mode. This condition, commonly referred to as the phase-matching condition, can generally be expressed as:
\begin{equation}
\label{eq:4}
    \left|\kappa_{xm}\hat{\mathbf x}+\kappa_{yn}\hat{\mathbf y}\right|=n_{\mathrm{eff}}^{\mathrm{p}}\kappa_0,
\end{equation}
where, the diffraction orders ($m,n$) represent the characteristics of the top grating, and $\kappa_{xm}$, $\kappa_{yn}$ are the wave numbers of the light diffracted along the $x$- and $y$- axes respectively.
In Fig.~\ref{fig1}, the illustration depicts the normal incidence of light onto the GMR from air, where the $xz$-plane is designated as the plane of incidence (POI). The incident light is considered to be $s$-polarized, with the electric field vector ($\mathbf E$-field) parallel to the $y$-axis as illustrated in Fig.~\ref{fig1}(b). For normal incidence of a harmonic wave, Eq.~\eqref{eq:4} can be rewritten as
\begin{equation}
\label{eq:7}
\sqrt{\left(\frac{2\pi m}{d_x}\right)^2+\left(\frac{2\pi n}{d_y}\right)^2}=\pm \kappa_0 n_{\mathrm{eff}}^{\mathrm{p}}.
\end{equation}
The TE$_{0}$ waveguide modes propagating along the $\pm x$-directions are excited for $(m,n)=(\pm 1,0)$. The corresponding spectral peak position of the resonance, $\lambda_{0}^{\mathrm{TE}_0}=787.83$ nm, can be determined using $n^{\mathrm{TE}_0}_{\mathrm{eff}}=1.76$ in Eq.~\eqref{eq:7}. Similarly, $(m,n)=(0,\pm 1)$ diffraction orders excite TM$_{0}$ waveguide modes at $\lambda_{0}^{\mathrm{TM}_0}=744.55 $ nm,  propagating along $\pm y$-directions.
Equation~\eqref{eq:7} illustrates the influence of various parameters, such as wavelength, grating period, and refractive index, on the resonance of a GMRF. The direct relationship between the waveguide mode index $n_{\mathrm{eff}}^{\mathrm{p}}$ and the refractive index of the waveguide material $n_{\mathrm wg}$ indicates that the grating resonance can be adjusted by a minor alteration in the refractive index of the waveguide material. 
Nonlinear optical modulation of the GMR arises due to an intensity-dependent change in the refractive index. For a medium, which is isotropic in the linear regime, the Kerr effect at frequency $\omega$ can be described by \cite{boyd2008}
\begin{equation}
  P_i^{\rm NL} =\sum_j\epsilon_0\chi_{ij}^{(\rm Kerr)} E_j.\label{PE}
\end{equation}

Here, $P_i^{\rm NL}$ and $E_j$ denote the components of the electric polarization and electric field, respectively, with subscripts labeling the axes of the Cartesian laboratory frame. $\chi_{ij}^{(\rm Kerr)}$ represents the effective susceptibility tensor defined by
\begin{align}
  \chi_{ij}^{(\rm Kerr)} &= \left[n_{\mathrm{wg}}^2-1+A'|\mathbf{E}|^2\right]\delta_{ij}+B'\Re(E_iE_j^*),\label{chieff}
\end{align}
which also includes the linear contribution. In Eq.~\eqref{chieff}, $\Re$ denotes the real part and
\begin{align}
  A'&=6\chi_{1122}-3\chi_{1221},\nonumber\\
  B'&=6\chi_{1222} ,
\end{align}
where $\chi_{ijkl}\equiv\chi_{ijkl}^{(3)}(\omega=\omega+\omega-\omega)$ are the components of the third-order susceptibility tensor. Here, we choose the labeling of the directions such that 1 stands for the $x$ direction, and 2 and 3 for the $y$ and $z$, respectively. The effective permittivity tensor $\epsilon_{ij}^{(\rm Kerr)}$ is related with $\chi_{ij}^{(\rm Kerr)} $ by
\begin{align}
  \epsilon_{ij}^{(\rm Kerr)} &= \delta_{ij}+\chi_{ij}^{(\rm Kerr)}/\epsilon_0\label{epseff},
\end{align}
and the anisotropic effective refractive index $n^{\mathrm{(Kerr)}}_{\mathrm{wg},ij}$ can be evaluated from $n^{\mathrm{(Kerr)}}_{\mathrm{wg},ij}=\sqrt{\epsilon_{ij}^{(\rm Kerr)}}$. Equation~\eqref{epseff} suggests that, despite the isotropic nature of the material in the linear regime, the nonlinear interaction between light and matter induces an effect resembling anisotropic behavior in a linear optical medium. Specifically, in the scenario of a self-defocusing optical Kerr effect, $n_{\mathrm {wg}}(I_1)<n_{\mathrm {wg}}(I_0)$ for $I_1>I_0$, indicating a decrease in the refractive index of the material with increasing intensity of the incident light beam. Consequently, this phenomenon results in a blue shift of the resonance peak position.

It is crucial to emphasize that, for accurate estimation of the resonance peak shift, the modal field within the waveguide layer must be precisely determined through a rigorous solution of the diffraction problem for the GMR. Additionally, within a specific range of incident light intensities, the GMR may exhibit a spectral bistable response, necessitating further investigation.

Equations~\eqref{chieff} and \eqref{epseff} are valid for any arbitrary state of polarization of the incident electromagnetic (EM) wave, which differs from the case of linearly polarized light, where the Optical Kerr Effect can be sufficiently described by
\begin{equation}
n^{\mathrm{(Kerr)}}_{\mathrm{wg}}=n_{\mathrm{wg}}+ n^{\mathrm{nl}}_{\mathrm{wg}}I.
\label{neff}
\end{equation}
Here, $n^{\mathrm{nl}}_{\mathrm{wg}}$ represents the nonlinear refractive index of the material, capable of assuming either positive (self-focusing type OKE) or negative value (self-defocusing type OKE), $I$ is the intensity of light.
\subsection*{Full-wave numerical simulations and experimental results}
The GMR grating has been fabricated by creating an array of square-shaped SiN$_x$ pillars arranged in a square lattice on a fused silica substrate. The grating periods are $d_x=d_y=451$ nm and the grating fill factors of $f_x=f_y=0.46$ have been achieved in the fabricated structure. Figure~\ref{fig1}(c) presents a scanning electron micrograph (SEM) top-view of the grating. The wavelength-dependent refractive index ($n_{\mathrm{wg}}$) of SiN$_x$ is measured using ellipsometry (see section S3 of the supplementary information).

To experimentally demonstrate the nonlinear optical modulation of the GMR, we selected the shorter wavelength resonance associated with the TM$_0$ mode excitation in the guiding layer. The peak wavelength of the pulses was tuned close to the resonance wavelength  $\lambda_1=744.6$ nm using an optical parametric amplifier (OPA) unit. The pulse duration ($\tau$) was measured using a commercial GRENOUILLE device \cite{akturk2003}. The spectral and temporal characteristics of the pulses are detailed in Section S2 of the supplementary information.  Fig.~\ref{fig2}(a) shows the schematic of the experimental setup for transmission measurements with femtosecond laser pulses. 

The diffraction efficiencies of the GMR in direct transmission are plotted against the wavelength in air ($\lambda_0$) for two different average powers of the incident light, $P_1=0.5$ mW (solid lines), and $P_1=4.5$ mW (dashed lines). The plots in `Black' color represent numerical simulation results, while the `Red' lines represent experimental results in Fig.~\ref{fig2}(b). The third-order susceptibility $\chi^{\mathrm{(3)}}$ of SiN$_x$ is measured with a custom built $z$-scan set-up. To get adequate nonlinear signal in the $z$-scan measurements, a thicker SiN$_x$ film on a fused silica substrate was used. This thicker film possesses identical optical properties to the one used in the grating fabrication, as both were synthesized under identical deposition conditions. Numerical simulations are conducted using the Fourier Modal Method (FMM) for gratings made with materials exhibiting Kerr-type nonlinearity \cite{bej2014}. $\chi^{\mathrm{(3)}}$ of SiN$_x$ measured with $z$-scan is used in the nonlinear FMM simulations. Moreover, we assume that the origin of the nonlinearity in SiN$_x$ is the nonlinear response of bound electrons to the applied electromagnetic field. Hence, $B^{\prime}/A^{\prime}=2$ is used in Eq.~\eqref{epseff} \cite{boyd2008}. The complete set of numerical simulation and experimental results is included in Section S5 of the supplementary information. We observe a reduction in the modulation depth of the resonances in the experimental results compared to the numerical simulations. This discrepancy can be attributed to the finite size and reduced spatial coherence of the beam. For the results obtained with higher powers of the laser beam, the two-photon absorption of the SiN$_x$ material also reduces the modulation depth due to non-negligible nonlinear optical losses. The two-photon absorption coefficient $\beta^{\mathrm{TPA}}$ is also measured in $z$-scan. The experimental and the fitted $z$-scan results are included in section S5 of the supplementary information. The original and the shifted resonance peak positions $\lambda_1$, and $\lambda_2$ are marked by the `Green' and the `Blue' circles in Fig.~\ref{fig2}(b). Examining at $\lambda_2$, a relative change of over 50\% in diffraction efficiency is observed due to the nonlinear optical modulation of the GMR, effectively switching the GMR grating from a transmission ‘On’ to ‘Off’ state.

The peak positions of the resonances are plotted against the average power of the pulsed laser input ($P_{\mathrm{avg}}$) in Fig.\ref{fig2}(c). The lowest power used in the measurements was kept in the linear regime at $P_{\mathrm{avg}}=0.5$ mW. Both in the numerical simulations and experiments, we observe blue shifts of the resonances, resulting from the reduction of the effective index of the TM$_0$ mode as can be understood from Eq.\eqref{eq:7}. Noticeable differences in the resonance peak positions between the numerical and experimental results are observed. These differences are due to the fabrication errors. To compare the Kerr nonlinearity-induced resonance peak shifts in the numerical simulations and the experiments, we also plot the relative (with respect to the case with $P_{\mathrm{avg}}=0.5$ mW) peak shifts $\Delta\lambda_{\mathrm{peak}}$ in Fig.~\ref{fig2}(d). The error bars in the measurement results stem from uncertainties in measurements associated with fluctuations in the pulse peak power, which impact the incident light intensity on the grating. We observe excellent agreement between the numerical and the experimental results in  Fig.~\ref{fig2}(d).
\begin{figure}[H]%
\centering
\includegraphics[width=\textwidth]{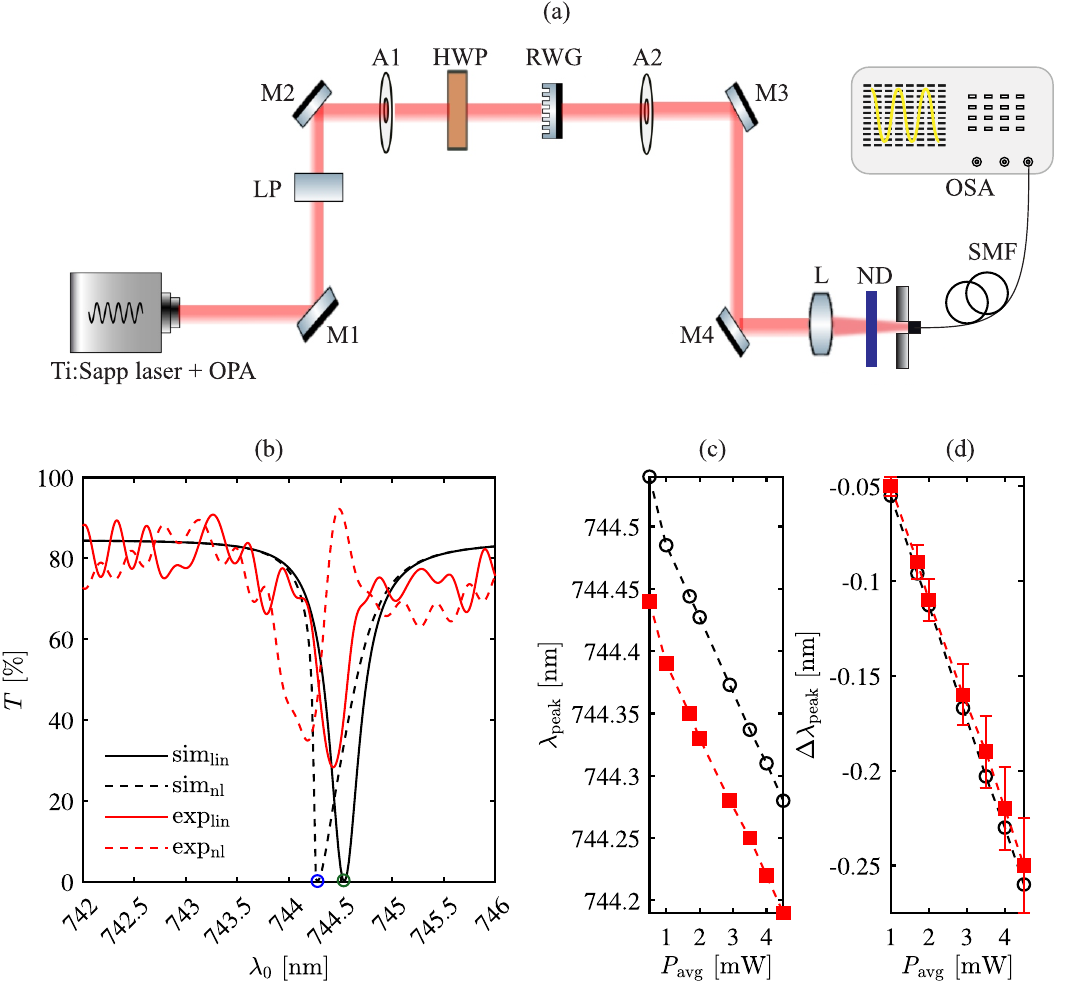}
\caption{(a) Schematic of the experimental set-up. SMF is a single-mode fiber, L is a lens, ND is a variable neutral density (ND) filter, LP is a linear polarizer, M1-M4 are mirrors, A1-A2 are apertures, and HWP is a half-wave plate. (b) Transmittance for normal incidence. Black and red lines represent numerical simulations and experimental results, respectively. The plots with solid lines are obtained for a small average power ($P_{\mathrm{avg}}=P_1$=0.5 mW) of the pulses. The plots with the dotted lines are for $P_{\mathrm{avg}}=P_2$=4.5 mW. The resonance minima for $P_1$, and $P_2$ are highlighted with the green and the blue circles respectively. (c) peak positions on the wavelength axis plotted against the average power ($P_{\mathrm{avg}}$) of the pulses. The red squares, and the black circles correspond to experimental and numerical simulation results respectively. (d) Relative resonance peak shifts calculated with respect to the reference peak position corresponding to the linear limit (marked by green circle in figure (b)). The error bars are calculated assuming 10 \% fluctuation in $P_{\mathrm{avg}}$.}\label{fig2}
\end{figure}

\subsection*{Electric field and permittivity distributions}
\begin{figure}[H]%
\centering
\includegraphics[width=\textwidth]{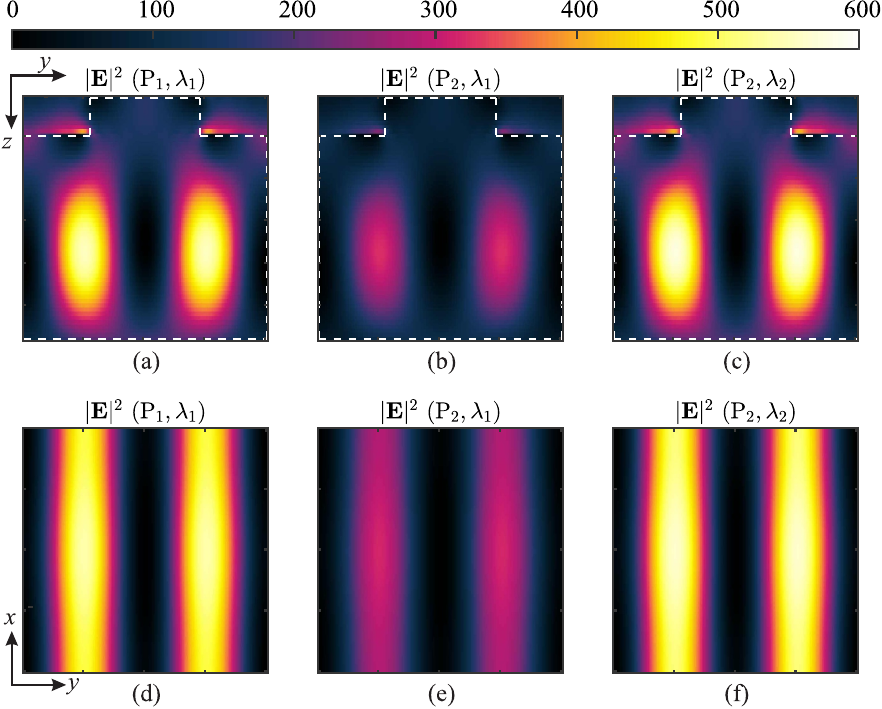}
\caption{Electric field intensity distributions inside the GMR grating over one grating period. (a)-(c) Field intensity maps in the $yz$ cross-sectional planes in the center of the structure, and (d)-(f) in the $xy$ cross-sectional planes in the middle of the waveguide layer.}\label{fig3} 
\end{figure}

To gain a better understanding of the nonlinear optical modulation dynamics of the GMR grating, we first visualize the electric fields inside the grating. Figures~\ref{fig3}(a)-(c) illustrate the fields inside the GMR in $yz$- cross-sectional planes taken at $x=0$ (middle of the square-shaped pillar) for $y$- polarized light pulses. The color maps are scaled between `0' and `600' for the best visualization, providing a clear representation of the field distribution. The white dashed lines mark one grating period. The result in Fig.~\ref{fig3}(a) is obtained with the laser peak wavelength at $\lambda_0=\lambda_1$ (marked by the green circle in Fig.~\ref{fig2}(a)), and for pulse peak power $P_{\mathrm{avg}}=P_1=0.5$ mW. Similar plot is shown in Fig.~\ref{fig3}(b) for $P_{\mathrm{avg}}=P_2=4.5$ mW. Compared to the field in Fig.~\ref{fig3}(a), we notice a weaker field confinement in Fig.~\ref{fig3}(b). This can be explained in terms of the weaker resonance magnitude at $\lambda_1$ for higher power ($P_2$) of the pump pulse. By examining the field distribution at $\lambda_2$, i.e., at the shifted resonance peak wavelength, we notice the full recovery of the field confinement in Fig.~\ref{fig3}(c). Electric fields in the $xy$- cross-sectional planes, with $z$ coordinate set at the middle of the waveguide layer, are plotted in Figs.~\ref{fig3}(d)-(f). The $xy$- view provides insight into the field distribution and mode propagation direction within the waveguide layer. The plots show that the TM$_0$ modes propagate along $\pm y$- directions.

\begin{figure}[H]%
\centering
\includegraphics[width=\textwidth]{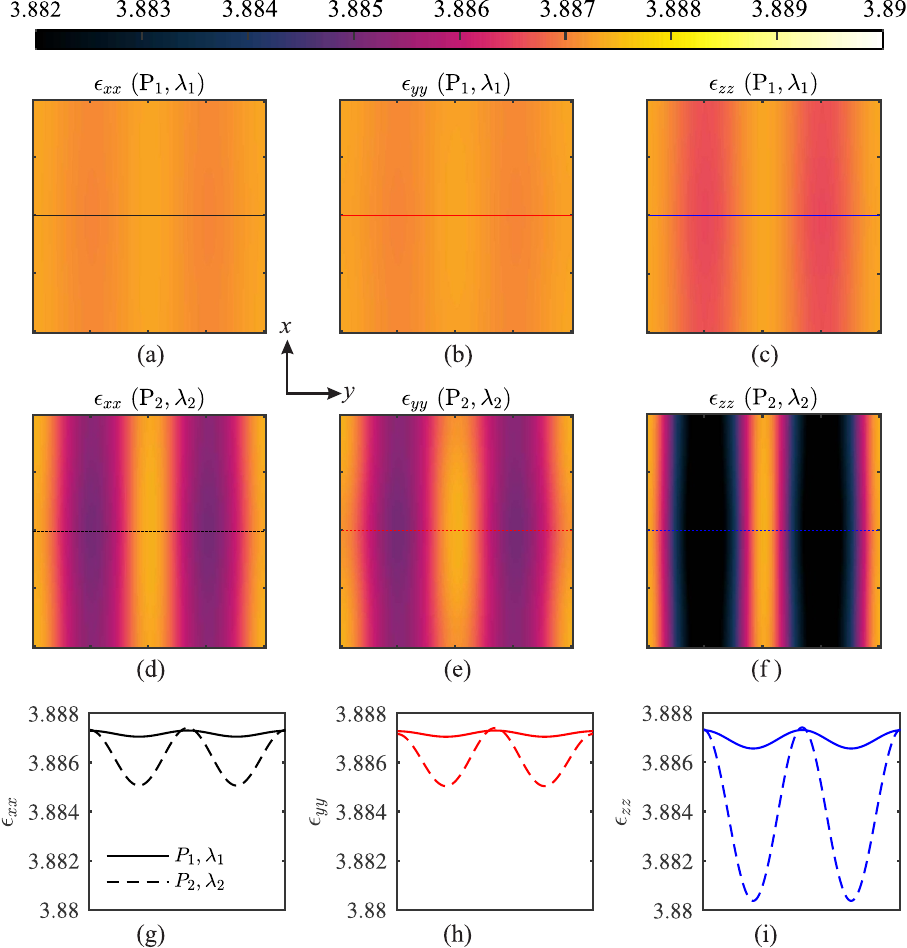}
\caption{Distribution of the relative permittivity tensor components $\epsilon_{ij}^{(\rm {Kerr})}$ in $xy$- cross-sectional plane through the middle
of the waveguide layer. Maps of the tensor component $\epsilon_{xx}$ in (a) and (d), $\epsilon_{yy}$ in (b) and (e), and $\epsilon_{zz}$ in (c) and (f) over one grating period are calculated. (g)-(i) Line scans  of the permittivity maps in (a)-(f).}\label{fig4}
\end{figure}

To illustrate the onset of the guided-mode enhanced Kerr effect, and light-induced anisotropy, we plot permittivity maps in $xy$- cross-sectional plane at the middle of the waveguide layer. Figures~\ref{fig4}(a)-(f) depict the permittivity distributions inside one period of the grating obtained using the Fourier Modal Method (FMM) approach \cite{bej2014}. In Figs.~\ref{fig4}(a)-(c), the diagonal components of the effective permittivity tensor $\mathbf{\epsilon}_{ij}^{\mathrm{(Kerr)}}$, i.e., $\epsilon_{xx}$, $\epsilon_{yy}$, and $\epsilon_{zz}$ inside one grating period are plotted for the average power $P_1=0.5$ mW, and wavelength $\lambda_1$ of the light incident on the GMR. We observe very small changes in permittivities in Figs.~\ref{fig4}(a)-(c). Moreover, relatively stronger changes are observed in $\epsilon_{zz}$ (Fig.~\ref{fig4}(c)) as compared to $\epsilon_{xx}$, and $\epsilon_{yy}$ (Figs.~\ref{fig4}(a)-(b)). This can be explained by the relatively stronger $z$-component of the $\mathbf{E}$-field at resonance as compared to the $x$, and $y$ components, which makes stronger contribution to the second term in Eq.~\eqref{chieff} for $\epsilon_{zz}$. Additionally, it is noteworthy that $\epsilon_{xx}=\epsilon_{yy}$ due to the symmetry of the fields in this case of a square-shaped structure and the symmetry in the light-matter interaction (normal incidence). In Figs.~\ref{fig4}(d)-(f), maps of $\epsilon_{xx}$, $\epsilon_{yy}$, and $\epsilon_{zz}$ inside one grating period are plotted at the shifted resonance peak wavelength $\lambda_2$ for an average power of $P_1=4.5$ mW. Variations of $\epsilon_{xx}$, $\epsilon_{yy}$, and $\epsilon_{zz}$ along the lines drawn through middle of one grating period (as shown in Figs.~\ref{fig4}(a)-(f)) for the two different cases with different average powers and wavelengths of excitation are shown in Fig.~\ref{fig4}(g), (h), and (i) respectively. The figures reveal a three-order-of-magnitude maximum change in the permittivity values. Furthermore, the plots in Figs.~\ref{fig4}(a)-(i) demonstrate the onset of light-induced anisotropy inside the grating.

\subsection*{Dynamic pulse shaping}
\begin{figure}[H]%
\centering
\includegraphics[width=\textwidth]{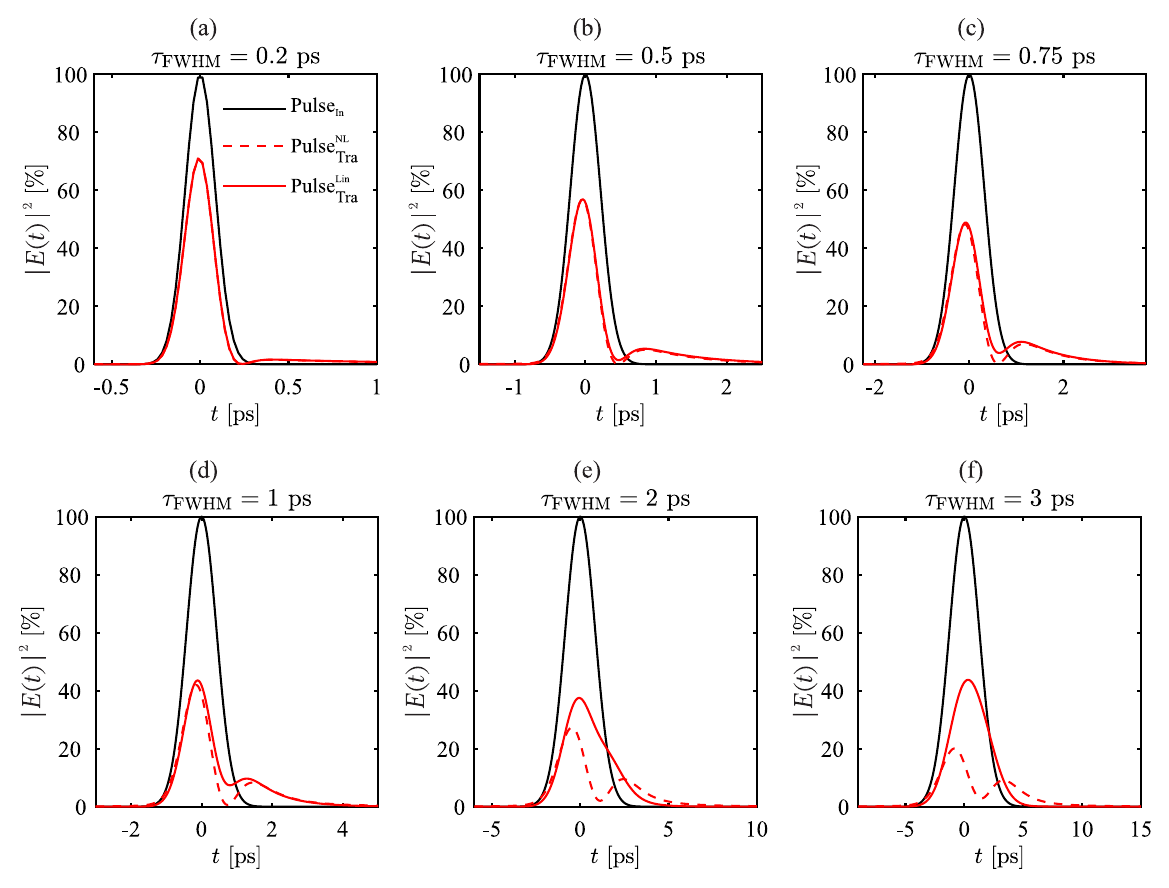}
\caption{
Dynamic shaping of time-domain optical pulses achieved by harnessing the GMR-enhanced Optical Kerr Effect (OKE). Figures (a)-(f) depict plots illustrating the incident (solid black line) and transmitted (solid and dashed red lines) electric field intensities of the pulses across a range of pulse durations from 0.2 ps (a) to 3 ps (f). The solid red and dotted red lines correspond to average pulse powers $P_1$ and $P_2$, respectively.} \label{fig5}
\end{figure}

We explore the capability of the GMR device for shaping time-domain light pulses by harnessing the amplified optical Kerr effect within the grating. The numerical simulations span a range of pulse durations from 0.2 ps to 3 ps, as illustrated in Figs.~\ref{fig5}(a)-(f). The transmitted pulses are analyzed for the grating responses in both linear and nonlinear regimes, assuming that the pulse peak power remains the same for each pulse duration. The underlying mathematics behind pulse shaping can be conveniently expressed in the frequency domain as
\begin{equation}
E_{\mathrm{out}}(\omega)=\eta_{\mathrm{GMR}}(\omega)\cdot E_{\mathrm{in}}(\omega).
\label{pulse_shaping_2}
\end{equation}
Here, $E_{\mathrm{in}}$, and $E_{\mathrm{out}}$ represent the $\mathbf{E}$-fields associated with the incident and the transmitted pulses respectively. $\eta_{\mathrm{GMR}}$ denotes the grating response in the frequency domain.
The time-domain plots presented in Figs.~\ref{fig5}(a)-(f) are obtained using Eq.~\eqref{pulse_shaping_2} under the assumption that the peak frequency of the incident pulses is $\omega_2=2\pi c/\lambda_2$. The solid black lines represent the incident pulses, while the solid and dashed red lines correspond to the transmitted pulses at two different peak powers, $P^{\mathrm p}_1=2.9$ MW and $P^{\mathrm p}_2=26.77$ MW, respectively. These peak powers correspond to the average pulse powers of $P_1=0.5$ mW, and $P_2=4.5$ mW, respectively for pulses of duration $\tau=160$ fs. 
It is noteworthy that for the shortest pulse duration of $\tau=0.2$ ps, the self-phase modulation effects induced by the GMR-enhanced Kerr nonlinearity do not cause a significant difference between the transmitted pulses for the two different peak powers of the incident pulses. This is because the pulse spectrum is much broader compared to the resonance spectrum. Consequently, the nonlinearity-induced changes affect only a narrow spectral window of the incident pulse. However, as the pulse duration increases, leading to a narrower pulse spectrum, noticeable distinctions emerge between the transmitted pulses for average powers of $P_1$ and $P_2$. Specifically, for $\tau=3$ ps, the maximum disparity is observed, where the GMR-enhanced self-phase modulation nearly induces pulse splitting.

\section*{Summary}\label{sec3}

In summary, we present the first experimental demonstration of ultrafast nonlinear optical modulation of guided-mode resonance in a diffraction grating. The grating, fabricated using a CMOS-compatible, low-loss SiN$_x$ material, shows an enhancement in the Optical Kerr Effect (OKE) by more than three orders of magnitude compared to a bare film of the same thickness. Our experimental results reveal over 50\% modulation of the resonance magnitude across the spectrum, attributed to Kerr nonlinearity-induced shifts in the resonance peak, enabling all-optical switching between 'On' and 'Off' transmission states. These findings are validated through comprehensive numerical simulations and a semi-analytical approach for qualitative understanding. Visualization of the field and permittivity distributions within the grating further clarifies the origin of the GMR-enhanced OKE, resulting in strong and tunable light-induced anisotropy. Additionally, we demonstrate the potential of GMR-enhanced OKE for all-optically tunable pulse shaping.

\section*{Discussion}\label{sec4}

We observed strong agreement between the magnitude of the optical Kerr effect, indicated by the spectral shift of the resonance dip, and the results from $z$-scan measurements of a bare SiN$_x$ film. This finding highlights the potential of using guided-mode resonance (GMR)-enhanced Optical Kerr Effect (OKE) as an alternative method for nonlinear optical material characterization. Unlike traditional methods, such as the $z$-scan technique that requires tightly focused pulsed light, which can cause localized heating or damage, the GMR-enhanced OKE approach allows for measurements with a collimated beam, significantly reducing the risk of material damage.

This research opens new avenues for exploiting Kerr-type nonlinearities in all-dielectric gratings, paving the way for advancements in high-resolution nonlinear imaging, where enhanced responses improve contrast and sensitivity, and in laser engineering, where precise light-matter interaction is key. The findings also have important implications for all-optical signal processing, which demands fast, tunable, low-power components. The ability to dynamically modulate light using all-dielectric materials without high losses offers a promising direction for developing next-generation photonic devices and systems.

\section*{Methods}\label{sec5}
\subsection*{Numerical simulations}
Numerical simulations are performed using an in-house developed program based on the Fourier Modal Method (FMM) for Kerr-type nonlinear materials in gratings. The experimentally determined $\chi^{\mathrm{(3)}}$ value is incorporated, considering only the nonlinear response of bound electrons for OKE. The permittivity distributions within a grating period are computed by iteratively evaluating electric-field components in a 3D grid of $n_x \times n_y \times n_z$ points using Fast Fourier Transform (FFT). The effective susceptibility in each grid is calculated based on the solved field components and updated iteratively until numerical convergence is reached. To demonstrate dynamic pulse shaping, the resonance spectra of the grating are multiplied with the incident pulse in the frequency domain, and the final results in the time domain are obtained through inverse Fourier transformation.

\subsection*{Fabrication}
The GMR device is fabricated using electron beam patterning and reactive ion etching of a SiNx film deposited on a 1 mm thick fused silica substrate. At first, the substrate is cleaned with acetone and isopropanol, followed by rinsing with DI water, and drying with nitrogen gun. After that, the top surface of the substrate is cleaned with oxygen plasma, and then coated with a 290 nm (±10 \%) SiN$_\mathrm{x}$ film using plasma-enhanced chemical vapor deposition with a gas mixture of 2 \% SiH$_4$/N$_2$:NH$_3$=100:3. The temperature and the pressure of the CVD chamber were fixed at 300°C and 1000 mTorr, respectively. A 50 nm Cr layer is then deposited on the SiN$_\mathrm{x}$ film using e-beam assisted evaporation, followed by spin-coating a positive electron beam resist AR-P 6200 (approximately 220 nm thickness). The e-beam resist was baked at 150°C for 3 minutes for solvent removal. Next, the grating grooves are patterned using a Raith electron-beam lithography instrument EBPG-5000+.  Ethyl 3-ethoxypropionate is used for resist development post patterning. The development process involves spinning the substrate at 4000 r.p.m. for 60 seconds. After the resist development, Cr is etched using a gas mixture of Cl$_2$ (54 sccm) and O$_2$ (4 sccm) for 3 minutes 30 seconds in Plasmalab 100 with AR-P 6200 as the etch mask. The residual resist layer is then removed with O$_2$ cleaning for 30 seconds. SiNx etching is carried out in Plasmalab 80 with a gas mixture of CHF$_3$/O$_2$ = 45/10 sccm. After that, chromium is removed using a commercial wet etchant Etch18, followed by rinsing with DI water and drying with nitrogen gun. The fabrication flow chart is shown in Fig.~S2 of the supplementary information.

\subsection*{Experimental procedure}
The grating transmittance is characterized using a femtosecond pulsed laser at varying average powers, with peak wavelength set to 744.6 nm using an optical parametric amplifier (OPA). Time duration, spectral FWHM, and phase distribution over time/spectrum of the pulses is determined through a commercial GRENOUILLE device \cite{akturk2003}. Detailed spectral and temporal pulse characteristics are provided in Section S2 of the supplementary information. The beam from the OPA undergoes filtering, expansion, and collimation before estimating its $1/e^2$ diameter with a CCD camera. Light polarization is controlled using a linear polarizer (LP) and a half-wave plate (HWP). The grating is excited by light pulses with their $\mathbf{E}$-field polarized along the $y$-axis. The transmitted signal is routed to an optical spectrum analyzer (OSA) through a short focal length lens and single-mode optical fiber, following signal attenuation with a variable neutral density (ND) filter to mitigate any possible nonlinear optical effects arising from the fiber itself. The refractive index and thickness of the SiN$_x$ film deposited on a fused silica substrate are measured using a spectroscopic ellipsometer (Semilab SE-2000) with a spectral range of 190 to 2500 nm. More information on the ellipsometry measurement and fitting procedure can be found in section S3 of the supplementary information. $z$-scan measurements are conducted to assess the third-order nonlinear optical susceptibility ($\chi^{(3)}$) of SiN$_x$. A thicker SiN$_x$ film (1.2 microns) deposited under the same CVD recipe is used for the $z$-scan procedure. Further specifics about the details of `z-scan' procedure, and fitting methodology of the experimental data for extraction of $\chi^{\mathrm{(3)}}$ can be found in section S1 of the supplementary information.

\backmatter

\bmhead{Supplementary information}
Supporting content for the article is provided in a separate document titled 'Supplementary Information: Ultrafast Modulation of Guided-Mode Resonance in a Kerr-Nonlinear Silicon Nitride Grating.' The supplementary information includes detailed descriptions of the following: (1) the z-scan procedure, (2) pulse spectral and temporal profile measurements with a commercial GRENOUILLE device, (3) ellipsometry measurements, (4) detailed fabrication procedure for the grating, (5) complete grating transmission spectra for a range of pulse powers, (6) spectral dispersion of the guided modes supported by the grating, and (7) grating based dynamic pulse shaping simulations in both time and frequency domains.

\bmhead{Acknowledgments}
S. B. gratefully acknowledges the insightful discussions with the late Emeritus Prof. Jari Turunen, which led to the concept behind this project. The authors also extend their sincere thanks to Dr. Hannu Pasanen for his support in establishing the experimental setup, and to Dr. Petri Karvinen for his essential advice during the fabrication of the grating structure.

\section*{Declarations}

\begin{itemize}
\item {\bf{Funding}}- This work was supported by the Flagship of Photonics
Research and Innovation (PREIN) funded by the Academy of
Finland—Grant Nos. 31001498, and 346511.
\item {\bf{Conflict of interest}}- The authors declare no conflict of interest.
\item {\bf{Availability of data and materials}}- The data related to this research are available from the corresponding author upon reasonable request.
\item {\bf{Authors' contributions}}- S.B. conceived the idea behind this work, planned the research activities jointly with T.N. S.B. designed and fabricated the grating structure, established the experimental setup, performed the measurements, and analyzed the experimental data. T.N. supported the numerical simulations and the analysis of the experimental data. N.T. assisted in setting up the experiments and provided valuable guidance for the alignment of the laser and the OPA. T.N. and R.F. provided suggestions throughout the progress of the work. S.B. prepared the manuscript with inputs from T.N. S.B., T.N., and R.F. collaboratively edited the manuscript.
\end{itemize} 





\end{document}